\def\tipo{2}      
\ifnum \tipo = 2
\documentstyle[prl,times,twocolumn,aps,epsfig]{revtex} \def\figsize{8cm}
\def\figsiz1{7cm} \def\frontmatter{\twocolumn[\hsize\textwidth\columnwidth
\hsize\csname@twocolumnfalse\endcsname} \else
\documentstyle[preprint,aps,prl,epsfig]{revtex} \def\figsize{9cm}
\def\figsiz1{8cm} \def\frontmatter{}  \fi

\begin{document} 
\draft \frontmatter
\title{Characterization of Sleep Stages by Correlations in the
Magnitude and Sign of Heartbeat Increments}
\author{Jan W.~Kantelhardt$^{1}$, Yosef Ashkenazy$^{1}$,
Plamen Ch.~Ivanov$^{1,2}$, Armin Bunde$^{3}$, \\
Shlomo Havlin$^{4,1,3}$, Thomas Penzel$^{5}$,
J\"org-Hermann Peter$^{5}$, and H.~Eugene Stanley$^{1}$}
\address{$^{1}$ Center for Polymer Studies and Department of Physics,
Boston University, MA 02215, USA}
\address{$^{2}$ Beth Israel Deaconess Medical Center,
Harvard Medical School, Boston, MA 02215, USA}
\address{$^{3}$ Institut f\"ur Theoretische Physik III,
Justus-Liebig-Universit\"at, D-35392 Giessen, Germany}
\address{$^{4}$ Gonda-Goldschmied-Center and Department of Physics, 
Bar-Ilan University, Ramat-Gan 52900, Israel}
\address{$^{5}$ Klinik f\"ur Innere Medizin, Klinikum der 
Philipps-Universit\"at, D-35033 Marburg, Germany}
\maketitle
\begin{abstract}
We study correlation properties of the magnitude and the sign of the
increments in the time intervals between successive heartbeats during
light sleep, deep sleep, and REM sleep using the detrended fluctuation 
analysis method.  We find short-range anticorrelations in the 
sign time series, which are strong during deep sleep, weaker during
light sleep and even weaker during REM sleep.  In contrast, we find
long-range positive correlations in the magnitude time series, which
are strong during REM sleep and weaker during light sleep.  We observe
uncorrelated behavior for the magnitude during deep sleep.  Since the 
magnitude series relates to the nonlinear properties of the original
time series, while the signs series relates to the linear properties,
our findings suggest that the nonlinear properties of the heartbeat
dynamics are more pronounced during REM sleep.  Thus, the sign and the 
magnitude series provide information which is useful in distinguishing
between the sleep stages.
\end{abstract}
\ifnum \tipo = 2 ] \fi

\section{Introduction}

Healthy sleep consists of cycles of approximately 1--2 hours duration. 
Each cycle is characterized by a sequence of sleep stages usually 
starting with light sleep, followed by deep sleep, and rapid eye 
movement (REM) sleep \cite{REF1}.  While the specific functions of the 
different sleep stages are not yet well understood, many believe that
deep sleep is essential for physical rest, while REM sleep is important 
for memory consolidation \cite{REF1}.  It is known that changes in the
physiological processes are associated with circadian rhythms (wake or
sleep state) and with different sleep stages 
\cite{EPL-2,EPL-4,EPL-22,EPL-21}.  

Here we investigate how the heart rhythms of healthy subjects change 
within the different sleep stages.  Typically the differences in cardiac 
dynamics during wake or sleep state, and during different sleep stages are 
reflected in the average and standard deviation of the interbeat interval 
time series \cite{EPL-21,other}.  However, heartbeat dynamics
exhibit complex behavior which is also characterized by long-range 
power-law correlations \cite{Musha82,Peng93,Peng95}, and recent studies
show that changes in cardiac control due to circadian rhythms or different
sleep stages can lead to systematic changes in the correlation (scaling)
properties of the heartbeat dynamics.  In particular it was found that 
the long-range correlations in heartbeat dynamics change during wake and 
sleep periods \cite{EPL99}, indicating different regimes of intrinsic 
neuroautonomic regulation of the cardiac dynamics, which may switch on 
and off with the circadian rhythms.  Moreover, different sleep stages 
during nocturnal sleep were found to relate to a specific type of 
correlations in the heartbeat intervals \cite{PRL00}, suggesting a 
change in the mechanism of cardiac regulation in the process of sleep.

We employ a recently proposed approach of magnitude and sign analysis
\cite{PRL01,PRE02} to further investigate how the linear and nonlinear
properties of
heartbeat dynamics change during different stages of sleep.  We focus
on the correlations of the sign and the magnitude of the heartbeat 
increments $\delta\tau_i\equiv\tau_i-\tau_{i-1}$ obtained from 
recordings of interbeat intervals $\tau_i$ from healthy subjects 
during sleep (Fig.~1), where $i$ indexes each heartbeat interval.  
We apply the detrended fluctuation analysis (DFA) method on both
the sign and the magnitude time series.
We find that the sign series exhibits anticorrelated behavior
at short time scales which is characterized by a correlation exponent
with smallest value for deep sleep, larger value for light sleep, and
largest value for REM sleep.  The magnitude series, on the other
hand, exhibits uncorrelated behavior for deep sleep, and long-range
correlations are found for light and REM sleep, with a larger exponent
for REM sleep.  The observed increase in the values of both the
sign and magnitude correlation exponents from deep through light to 
REM sleep is systematic and significant.  We also find that the values of 
the sign and magnitude exponents for REM sleep are very close to the
values of these exponents for the wake state.

Recent studies suggest that (i) long-range correlated behavior of the
magnitude series obtained from a long-range anticorrelated increment series 
$\delta\tau_i$ relates to the nonlinear properties of the signal, while
the sign series reflects the linear properties \cite{PRL01,PRE02}, and (ii)
the increments in the heartbeat intervals are long-range anticorrelated
\cite{Peng93} and exhibit nonlinear properties
\cite{PRL01,PRE02,Nat96,Nat99-26,Nat99}.
Thus, our finding of positive power-law correlations for the magnitude 
of the heartbeat increments during REM sleep and of loss of these 
correlations during deep sleep indicates a different degree of 
nonlinearity in the cardiac dynamics associated with different sleep 
stages.  Our results may be useful, when combined with earlier studies 
of interbeat interval correlations in different sleep stages \cite{PRL00}, 
for distinguishing the different sleep stages using electrocardiogram
records.

The paper is organized as follows: In Section II we review the DFA method.
In Section III we apply the DFA to analyze the sign and magnitude time
series of healthy subjects.  In Section IV the significance of the results
and interpretations are discussed.

\section{Detrended fluctuation analysis}

In recent years the DFA method \cite{Peng94,SMDFA1,Peng95,PRL00} is
becoming a widely-used technique for the detection of long-range
correlations in noisy, nonstationary time series
\cite{taqqu,physa,kunhu,kunhu1,Peng95,EPL99,PRL00,PRL01,PRE02,buldy95,%
buldy98,neuron,bahar,gait,koscielny98,Ivanovameteo1999_12,talkner,cloud,%
malamudjstatlaninfer1999,Alados2000,economics,Liu,Vandewalle1999,fest1,fest2}.
It has successfully been applied to diverse fields such as DNA sequences
\cite{buldy95,buldy98}, heart rate dynamics \cite{Peng95,EPL99,PRL00,PRL01,PRE02},
neuron spiking \cite{neuron,bahar}, human gait \cite{gait}, long-time 
weather records \cite{koscielny98,Ivanovameteo1999_12,talkner}, cloud 
structure \cite{cloud}, geology \cite{malamudjstatlaninfer1999}, 
ethnology \cite{Alados2000}, economics time series 
\cite{economics,Liu,Vandewalle1999}, and solid state physics 
\cite{fest1,fest2}.  One reason we employ the DFA method is to avoid 
spurious detection of correlations that are artifacts of 
nonstationarities in the heartbeat time series.  Other techniques for 
the detection of correlations like the autocorrelation function and the 
power spectrum are not suited for nonstationary time series
(see e.~g. \cite{akselrod}).

The DFA procedure consists of four steps.

\noindent $\bullet$ {\it Step 1}: Determine the ``profile''
\begin{equation} \tilde{Y}(i) \equiv \sum_{k=1}^i x_k - \langle x \rangle, \qquad
i=1,\ldots,L \label{profile} \end{equation}
of the data series $x_k$ of length $L$.  Subtraction of the mean
$\langle x \rangle$ is not compulsory, since it would be eliminated by 
the later detrending in the third step.

\noindent $\bullet$ {\it Step 2}:
Divide the profile $\tilde{Y}(i)$ into $L_n \equiv [L/n]$ non-overlapping
segments of equal length $n$.  Since the length $L$ of the series is often
not a multiple of the considered time scale $n$, a short part at the end 
of the profile may remain.  In order not to disregard 
this part of the series, the same procedure is repeated starting from the
opposite end.  Thereby, $2 L_n$ segments are obtained altogether.

\noindent $\bullet$ {\it Step 3}: Calculate the local trend for each of
the $2 L_n$ segments by a least-square fit of the data.  Then we determine
the variance
\begin{equation} \tilde{F}^2_n(\nu) \equiv {1 \over n} \sum_{i=1}^{n}
[\tilde{Y}((\nu-1) n + i) - p_{\nu}(i)]^2 \label{fsdef}
\end{equation}
for each segment $\nu$, $\nu = 1, \ldots, 2 L_n$.  Here, $p_{\nu}(i)$ is 
the fitting polynomial in segment $\nu$.  Linear, quadratic, cubic,
or higher order polynomials can be used in the fitting procedure 
(conventionally called DFA1, DFA2, DFA3, $\ldots$) \cite{PRL00}.
Since the detrending of the time series is done by the subtraction of the 
polynomial fits from the profile, different order DFA differ in their 
capability of eliminating trends in the data.  In DFA$m$, $m$th order DFA,
trends of order $m$ in the profile (or, equivalently, of order $m - 1$ in the
original series) are eliminated.  Thus a comparison of the results for
different orders of DFA allows one to estimate the type of the polynomial
trend in the time series \cite{physa,kunhu}.  

\noindent $\bullet$ {\it Step 4:} Average over all segments and take the
square root to obtain the fluctuation function \cite{fn1},
\begin{equation} \tilde{F}(n) \equiv \left[ {1 \over 2 L_n} \sum_{\nu=1}^{2 L_n}
\tilde{F}^2_n(\nu) \right]^{1/2}. \label{fdef} \end{equation}
We are interested how $\tilde{F}(n)$ depends on the time scale $n$.  Hence,
we have to repeat steps 2 to 4 for several time scales $n$.  It is 
apparent that $\tilde{F}(n)$ will increase with increasing $n$.  If data $x_i$
are long-range power-law correlated, $\tilde{F}(n)$ increases, for large values
of $n$, as a power-law,
\begin{equation} \tilde{F}(n) \sim n^{\tilde{\alpha}} \label{alpha}. \end{equation}
For long-range correlated or anticorrelated data, random walk theory 
implies that the scaling behavior of $\tilde{F}(n)$ is related to the
autocorrelation function and the power spectrum.  If the time series is 
stationary, we can apply standard spectral analysis techniques and 
calculate the power spectrum $S(f)$ as a function of the frequency $f$.
Then, the exponent $\beta$ in the scaling law 
\begin{equation}
 S(f) \sim f^{-\beta}
\end{equation}
is related to the mean fluctuation function exponent $\tilde{\alpha}$ by
\begin{equation}
 \beta = 2\tilde{\alpha} - 1.
\end{equation}
If $0.5 < \tilde{\alpha} < 1$, the correlation exponent
\begin{equation}
\gamma = 2 - 2 \tilde{\alpha}
\end{equation}
describes the decay of the autocorrelation function
\begin{equation}
C(n) \equiv \langle x_i \, x_{i+n}\rangle \sim n^{-\gamma}. 
\end{equation}

We plot $\tilde{F}(n)$ as a function of $n$ on double logarithmic scales and
calculate $\tilde{\alpha}$ by a linear fit.  For uncorrelated data, the profile
$\tilde{Y}(i)$ corresponds to the profile of a random walk, and $\tilde{\alpha}
= 1/2$ corresponds to the behavior of the root-mean-square
displacement $R$ of the walk: $R(t) \sim t^{1/2}$, where $t$ is the time 
(number of steps the walker makes).  For short-range correlated data, a
crossover to $\tilde{\alpha} = 0.5$ is observed asymptotically for large scales
$n$.  If a power-law behavior with $\tilde{\alpha} < 0.5$ is observed, the
profile corresponds to anticorrelated fractional Brownian motion, and the
data $x_i$ are long-range anticorrelated (antipersistent).  Power-law 
behavior with $\tilde{\alpha} > 0.5$ indicates persistent fractional Brownian
motion, and the data $x_i$ are positively long-range correlated.  
In particular, for Gaussian distributed white noise with zero mean
(uncorrelated signal), we obtain $\tilde{\alpha} = 0.5$ from the DFA method.
In addition, DFA can be also used to determine the scaling exponent
for a wide variety of self-affine series, if the first step
[Eq.~(\ref{profile})] is skipped and the data are used directly instead
of the profile.  This way, the analysis is related to standard self-affine
and fractal analysis.

However, the DFA method cannot detect {\it negative} fluctuation
exponents $\tilde{\alpha}$, and it already becomes inaccurate for strongly
anticorrelated signals when $\tilde{\alpha}$ is close to zero.
Since strongly anticorrelated behavior (corresponding to $\tilde{\alpha}
\approx 0$) was previously reported for the heartbeat sign series,
we use a modified DFA technique \cite{PRL01}.  The most simple way to
analyze such data is to integrate the time series before the standard
DFA procedure.  Hence, we replace the {\it single} summation in
Eq.~(\ref{profile}), which is describing the determination of the
profile from the original data $x_k$, by a {\it double} summation,
\begin{equation} \label{double} Y(i) \equiv \sum_{k=1}^i
\left[ \tilde{Y}(k) - \langle \tilde{Y} \rangle \right]. \end{equation}
Following the DFA procedure as described above, we obtain a fluctuation
function $F(n)$ described by a scaling law as in Eq.~(\ref{alpha}), but
with an exponent $\alpha = \tilde{\alpha} + 1$,
\begin{equation} F(n) \sim n^{\alpha} \equiv n^{\tilde{\alpha}+1}.
\end{equation}
Thus, the scaling behavior can be accurately determined even if
$\tilde{\alpha}$ is smaller than zero (but larger than $-1$).  We note that
$F(n)/n$ corresponds to the conventional $\tilde{F}(n)$ in Eq.~(\ref{alpha}).
If we do not subtract the average values in each step of the summation in
Eq.~(\ref{double}), this summation leads to quadratic trends in the profile
$Y(i)$.  In this case we must employ at least the second order DFA to
eliminate these artificial trends.

\section{Correlation analysis of sign and magnitude time series}

To study the correlation properties of the sign $s_i\equiv{\rm sgn}
(\delta\tau_i)$ and magnitude $m_i\equiv\vert\delta\tau_i\vert$ 
obtained from the original interbeat increment time series $\delta\tau_i$, 
we investigate in parallel the corresponding double profiles [see
Eq.~(\ref{double})] for $x_i=s_i$ and $x_i=m_i$.
We calculate the fluctuation function $F(n)$ by DFA2 for a range
of time scales $4 \le n \le 200$.  The DFA2 method turned out to be
the most appropriate degree of detrending in an earlier study \cite{PRL00}.
In the following we will use the notation $\alpha_{\rm sign}$ for the
value $\alpha=\tilde{\alpha}+1$ of the sign series and $\alpha_{\rm mag}$
for the value $\alpha$ of the magnitude series.

We consider 24 records of interbeat intervals obtained from 12 healthy 
individuals during sleep.  The records have an approximate
duration of 7.5 hours.  Figure~1(a) shows the heartbeat interval time 
series for a typical healthy subject with periods of light sleep, deep 
sleep, REM sleep, and short intermediate wake phases.  The annotation 
and duration of the sleep stages were determined based on standard 
procedures \cite{rechtschaffen}.  Figure~1(b) shows a subset of the 
heartbeat interval series $\tau_i$, the increment series $\delta \tau_i$ 
as well as the corresponding series of sign $s_i$ and magnitude $m_i$.

In order to analyze the correlation properties during the different
sleep stages separately, we split each heartbeat interval series into
subsequences corresponding to the sleep stages.  Thus, from a typical 
7.5 hour series, we obtain several subsequences of heartbeat intervals
corresponding to light sleep, deep sleep, and REM sleep, as well as 
several subsequences corresponding to intermediate wake states.  In 
order to eliminate the effect of transitions between subsequent sleep 
stages, and because the determination of the sleep stages is done in 
intervals of 30 seconds, we disregard the first and last 50 seconds of 
each individual subsequence.  Then, to apply the DFA method, we calculate 
the profile for each subsequence (step 1), cut each of these profiles 
into segments (step 2), and calculate the variance for each segment 
(step 3).  In step 4, we calculate the fluctuation function $F(n)$ for
each sleep stage by averaging over all segments corresponding to (i) 
light sleep, (ii) deep sleep, and (iii) REM sleep, as well as for the
intermediate wake states. 

Figure 2 shows the normalized fluctuation functions $F(n)/n$
[corresponding to $\tilde{F}(n)$] versus the segment size (time
scale) $n$ for the sign and the magnitude for a representative subject.
We find short-range anticorrelated behavior for the sign of the 
increments [Fig.~2(a)].  Our analysis is performed for time scales $n\ge7$
beats --- just above the breathing peak \cite{fn2}.  In the intermediate
regime $7 \le n \le 20$ beats, we observe significant differences in 
the behavior of the fluctuation function $F(n)$ for the different sleep 
stages.  For deep sleep the $F(n)/n$ curve bends down characterized
by a correlation exponent $\alpha_{\rm sign} < 1$, for light
sleep the $F(n)/n$ curve remains flat ($\alpha_{\rm sign} \approx 1$),
and for REM sleep $F(n)/n$ increases with $n$ ($\alpha_{\rm sign} > 1$).
At $n \approx 20$, $F(n)/n$ exhibits a crossover, and at larger time scales
the observed anticorrelations slowly decay.  For $n > 100$ beats, we find 
uncorrelated behavior (characterized by $\alpha_{\rm sign} = 1.5$
for the profile after double integration) for all sleep stages.
For the sign series we also observe that the value of the fluctuation
function $F(n)$ at the position of the crossover ($n \approx 20$) is
significantly different for deep sleep, light sleep and REM sleep 
[Fig.~2(a)].  This observation, as well as the finding that a 
different correlation exponent $\alpha_{\rm sign}$ characterizes the 
behavior of $F(n)$ for different sleep stages in the intermediate regime 
$7 \le n \le 20$, could be of practical use in developing an algorithm 
that can automatically distinguish between different sleep stages based 
solely on heartbeat records.

In Fig.~2(b) we present our results of the DFA2 method for the magnitude
of the heartbeat increments.  In contrast to the short-range correlations
observed for the sign series, we find that the magnitude series for 
REM sleep is characterized by a scaling exponent $\alpha_{\rm mag} 
> 1.5$ for time scales $n > 10$, corresponding to positive long-range 
power-law correlations.  For light sleep, we find a smaller 
scaling exponent $\alpha_{\rm mag}$ than for REM sleep, indicating 
weaker long-range correlations.  Surprisingly we find that in contrast 
to REM and light sleep, the magnitude series for deep sleep is
uncorrelated, since the profile is characterized by
$\alpha_{\rm mag} = 1.5$ after the double integration.   This finding
is consistent with stronger multifractality during REM sleep than during
deep sleep \cite{mfneu}, since previous studies have related positive
long-range correlations in the magnitude series with multifractal and
nonlinear features present in the signal \cite{PRL01,PRE02}.
Following Refs.~\cite{Schreiber96,Schreiber00} we define a time series
to be linear if its scaling properties are {\it not} modified by
randomizing its Fourier phases.  In contrast, when applied to a nonlinear
series, the surrogate data test for nonlinearity, which is based on
Fourier phase randomization \cite{Schreiber96,Schreiber00}, generates a
linear series with different scaling properties for the magnitude series.

The nonlinearity of a time series is related to its multifractality.
The partition function $Z_q(n)$ of a time series, $x_i$, may be defined
as \cite{Bacry01},
\begin{equation} Z_q(n)=\langle |x_{i+n}-x_i|^q \rangle, \end{equation}
where $\langle \cdot \rangle$ denotes the average over the index $i$.
In some cases $Z_q(n)$ obeys scaling laws
\begin{equation} Z_q(n) \sim n^{\tau(q)}. \end{equation}
If the exponents $\tau(q)$ are linearly dependent on $q$ the series
$x_i$ is monofractal, otherwise $x_i$ is multifractal.  Monofractal
series fall under the category of linear series while multifractal 
series are classified as nonlinear series \cite{feder88}.  A possible
way to test this classification is to apply the surrogate data test
\cite{Schreiber96,Schreiber00}.  When this test is applied to a
multifractal series, it generates a linear series with a linear
dependence of $\tau(q)$ on $q$ in contrast to the nonlinear dependence
for the original series.  On the other hand, applying the surrogate
data test to a monofractal series does not affect its linear $\tau(q)$
dependence.

In \cite{PRL01,PRE02} it has been shown that the long-range correlations
in the magnitude series indicate nonlinear behavior.  Specifically, the
results suggested that the correlation exponent $\alpha_{\rm mag}$ of 
the magnitude series is a monotonically increasing function of the
multifractal spectrum width of the original series.  This conclusion
has been obtained based upon several examples of artificial multifractal
series \cite{PRL01,PRE02}.

Thus, the long-range magnitude correlations we find for REM sleep
indicate nonlinear contributions to the heartbeat regulation, which are
reduced during light and deep sleep.  Indeed, a multifractal analysis
of heartbeat intervals during daytime \cite{Nat99} indicated the presence
of multifractality.  In a recent study we also find stronger multifractality
during REM sleep than during deep sleep \cite{mfneu} which is consistent
with the scaling behavior of the magnitude series reported in the
present study.

\section{Significance of the results and summary}

The mean values and their standard deviations for the different sleep 
stages are shown in Fig.~3.  We estimate the exponents $\alpha$ from 
the slopes in the log-log plot of $F(n)$ versus $n$ for all records.  
Since the most significant differences for the short-range sign 
correlations occur in the range of $8 \le n \le 13$ heartbeats, we 
use this fitting range for the exponents $\alpha_{\rm sign}$. 
For the magnitude exponent $\alpha_{\rm mag}$, we use the range 
$11 \le n \le 150$, since the long-range correlations occurring in 
light and REM sleep can be observed best in this region.  We find 
that there is a significant difference in the sign series exponent
$\alpha_{\rm sign}$ observed for all three sleep stages (the $p$-value, 
obtained by the Student's $t$-test, is below $0.001$), and thus we
confirm the conclusions drawn from Fig.~2.  The magnitude correlation
exponents for REM sleep and for intermediate wake states are
significantly larger than those for the non-REM stages (light and deep
sleep).  Here also, the $p$-values are less than 0.001.  Note that
we do not find a significant difference between the average exponents 
for REM sleep and for the intermediate wake states.  This is not
surprising because heartbeat activity during REM sleep is very close to
heartbeat activity during the wake state and the heartbeat time series 
during REM and wake exhibit similar scaling properties \cite{EPL99,PRL00}.  

More significant than the differences for the
average exponents are the differences between the exponents for each
individual.  Figure~4 shows the $\alpha$ values for REM, light, and
deep sleep for all 12 healthy subjects (second night only).  In almost
all cases the exponent of the REM sleep is the largest, the
exponent of the light sleep is intermediate, and the exponent of
the deep sleep is smallest (there are three exceptions, indicated by
arrows). In our group of 24 records from 12 healthy individuals, we
find larger exponents in REM sleep than in deep sleep for $100 \%$ of
the sign series and for $88\%$ of the magnitude series. 

In a previous study of heartbeat records from healthy subjects during 
daytime activity, we found that the magnitude series is long-range 
correlated, while the sign series is short-range anticorrelated for all
subjects in the database \cite{PRL01,data}.  This finding suggests an
empirical ``rule'', namely that a large (small) heartbeat increment in 
the positive direction is most likely to be followed by a large (small)
increment in the negative direction, and that a large (small) increment
is most likely followed by large (small) increments.  Our present 
results suggest that this empirical ``rule'' also applies to REM sleep, 
while in deep sleep small and large increments seem to appear in a
random fashion.  On the other hand, the stronger sign anticorrelations 
in deep sleep indicate that a positive increment is more likely --- 
even more likely than in REM sleep --- to be followed by a negative 
increment.  Thus, the correlation behavior of the heartbeat increments 
and their signs and magnitudes during daytime activity is similar to 
the behavior we find in REM sleep, but quite different from the behavior 
we observe in deep sleep.  This is consistent with our finding
[Fig.~\ref{fig:3}] of average exponent values for the wake episodes 
similar to the exponent values for REM sleep.

In summary, we analyzed, for healthy subjects, interbeat interval
fluctuations during different sleep stages which are associated with 
different brain activity.  We find that the short-range anticorrelations 
in the sign of the increments are stronger during deep sleep, weaker 
during light sleep, and even weaker during REM sleep.  In contrast, the 
magnitude of the increments is long-range correlated with a larger 
exponent during REM sleep, suggesting stronger nonlinear contributions 
to the heartbeat dynamics in this stage compared with weaker nonlinear
contributions in the non-REM stages.

\section{Acknowledgements}

JK would like to thank the Minerva Foundation and the Deutscher
Akademischer Austauschdienst (DAAD) for financial support.  SH would
like to thank the Binational USA-Israel Science Foundation.  We also
would like to thank A. L. Goldberger for discussions, and the
NIH/National Center for Research Resources (P41 RR13622) for financial
support.  The healthy volunteers were recorded as part of the SIESTA
project funded by the European Union grant no. Biomed-2-BMH4-CT97-2040.

\begin{figure}\centering
\centerline{\epsfxsize\figsize\epsfbox{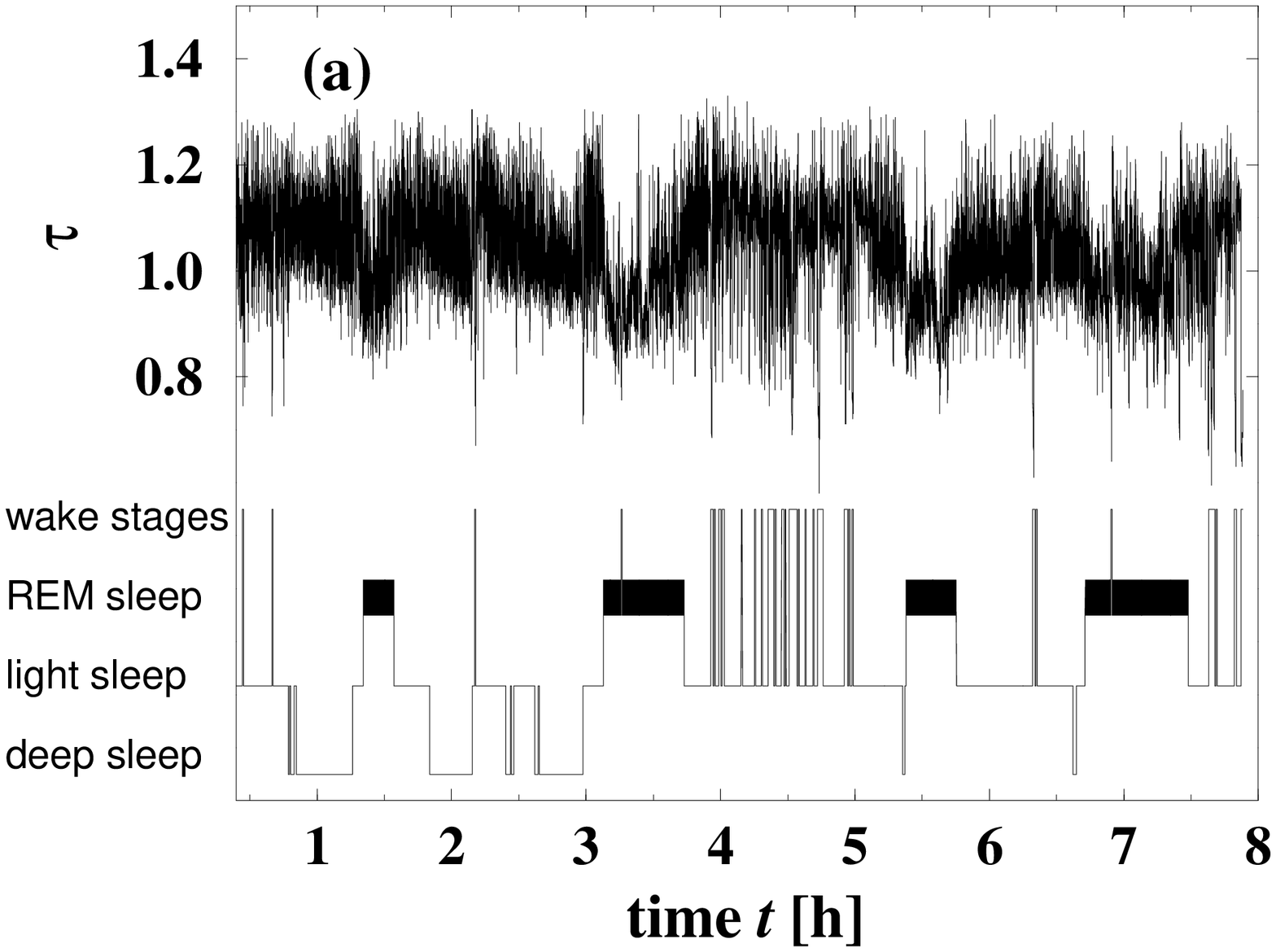}}
\centerline{\epsfxsize\figsize\epsfbox{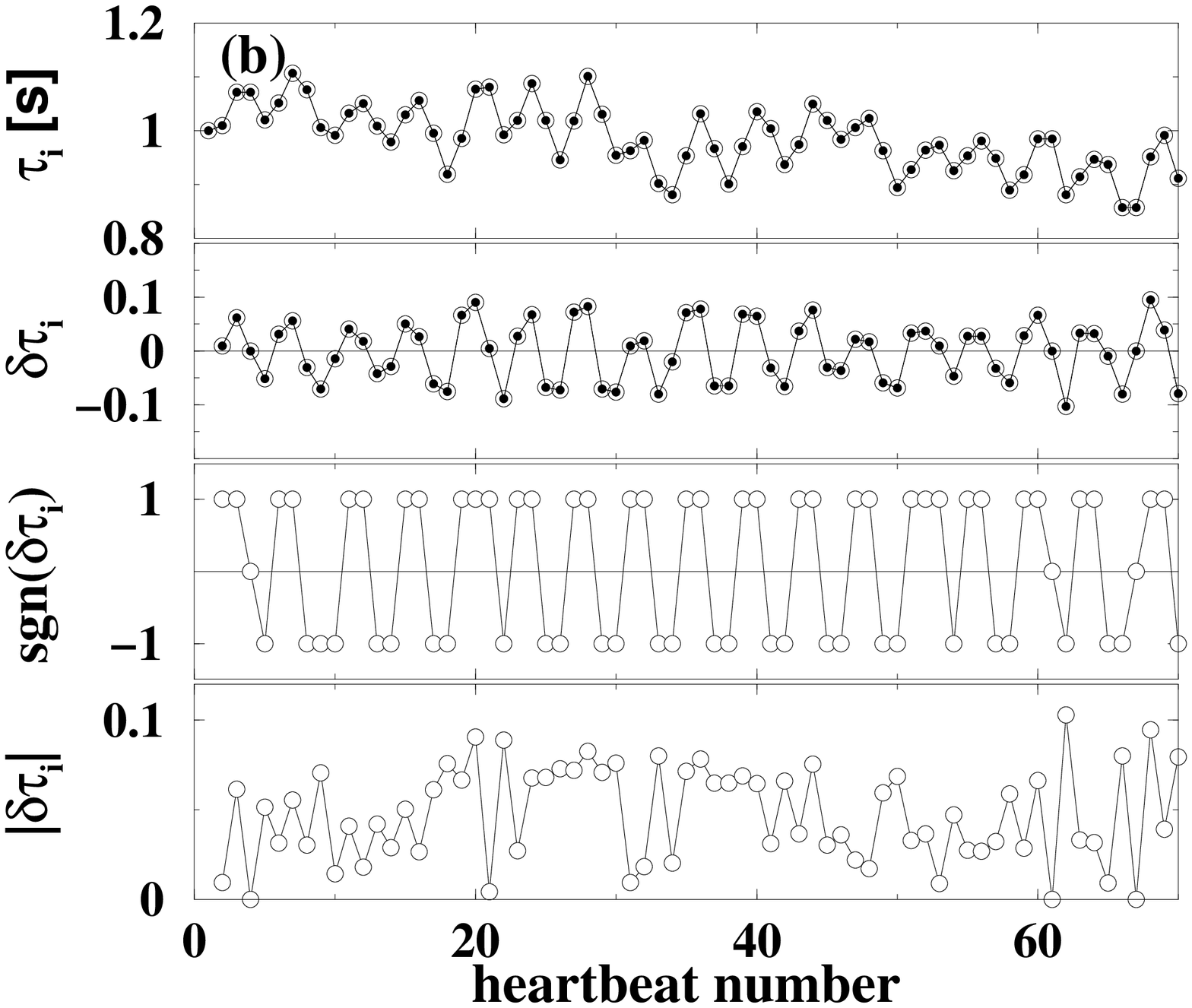}}
\vspace{1cm}
\parbox{\figsize}{\caption[]{\small
(a) One-night record for a healthy subject.  The intermediate wake states 
as well as REM sleep, light sleep, and deep sleep stages have been 
determined by visual evaluation of brain, eye and muscle activity 
\cite{rechtschaffen}.  (b) Heartbeat intervals $\tau_i$,
increments $\delta \tau_i\equiv \tau_i - \tau_{i-1}$, signs of the 
increments ${\rm sgn} (\delta \tau_i)$, and absolute increments $\vert 
\delta \tau_i \vert$ for a subset of the record shown in (a).}
\label{fig:1}}
\end{figure}

\begin{figure}\centering
\epsfxsize\figsize\epsfbox{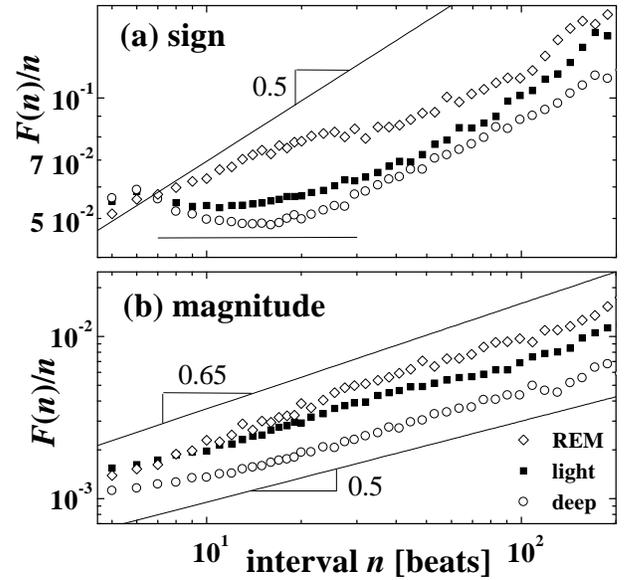}\vspace{0.5cm}
\parbox{\figsize}{\caption[]{\small
The normalized fluctuation functions $F(n)/n$ of the integrated
series of signs $s_i$ (a) and magnitudes $m_i$ (b) of the heartbeat
increments for a representative healthy subject.  $n$ is the time
scale in beat numbers.  Before applying the DFA2 the profiles have been 
split according to the sleep stages.  The fluctuation functions for the 
segments corresponding to the same type of sleep have been averaged 
with weights according to the number of intervals in each segment.
The different symbols correspond to the different sleep stages, light 
sleep, deep sleep, and REM sleep.}
\label{fig:2}}
\end{figure}

\begin{figure}\centering
\epsfxsize7.0cm\epsfbox{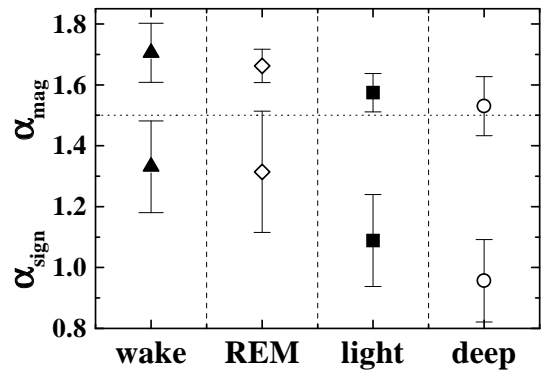}\vspace{0.5cm}
\parbox{\figsize}{\caption[]{\small
The average values of the fluctuation exponents $\alpha_{\rm mag}$ for
the magnitude series and $\alpha_{\rm sign}$ for the sign series for 
the different phases (wake state, REM sleep, light sleep, and deep sleep).   
For each of the 24 records from 12 healthy subjects the corresponding 2nd
order DFA fluctuation functions $F(n)$ have been fit by Eq.~(4)
in the range of $8 \le n \le 13$ and $11 \le n \le 150$ heartbeats for
$\alpha_{\rm sign}$ and $\alpha_{\rm mag}$, respectively, where
the most significant differences between the sleep stages occur. }
\label{fig:3}}
\end{figure}

\begin{figure}\centering
\epsfxsize\figsize\epsfbox{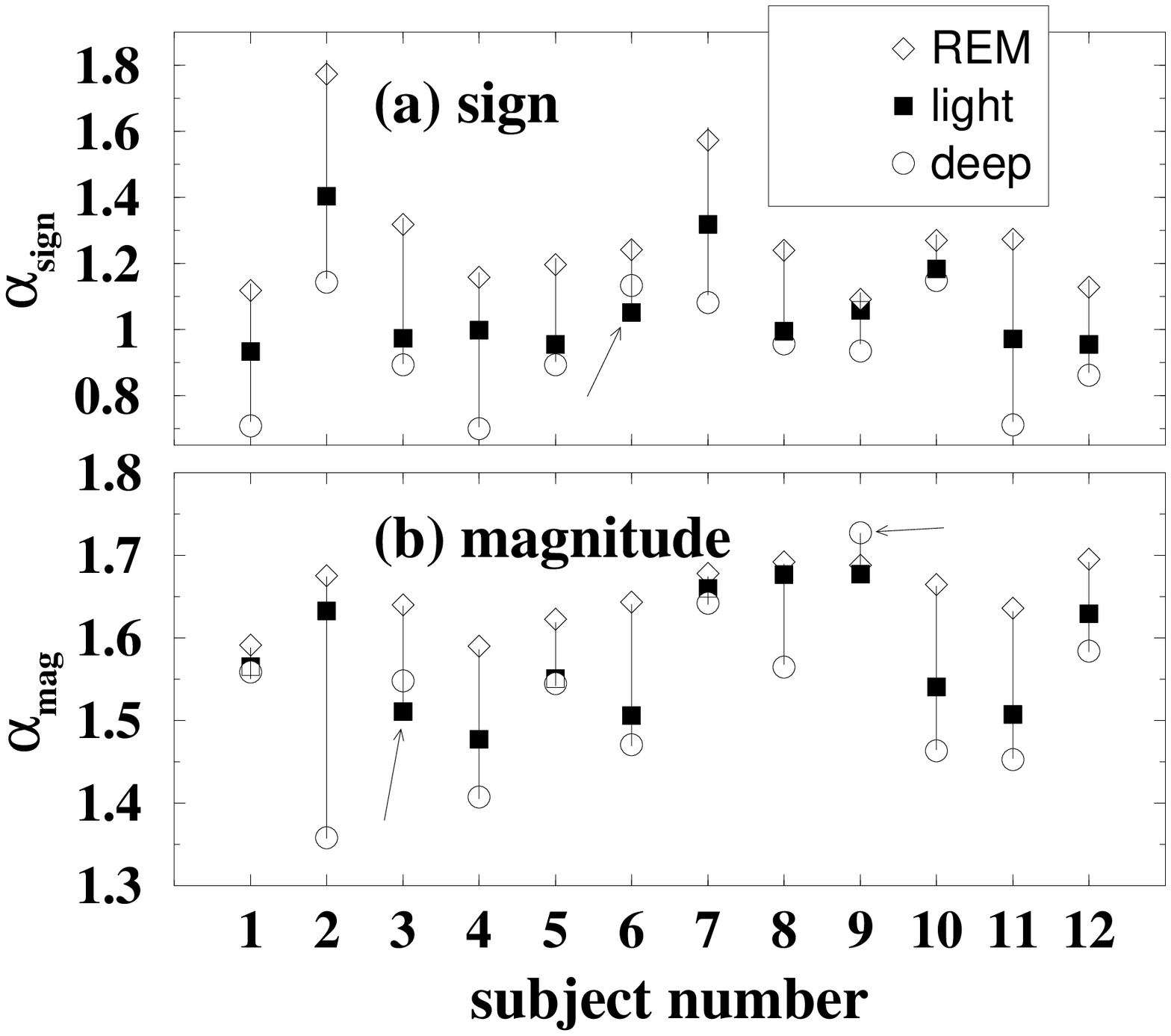}\vspace{0.5cm}
\parbox{\figsize}{\caption[]{\small The values of the effective
fluctuation exponents $\alpha$ for the integrated sign series (a)
and the integrated magnitude series (b) are shown for all 12 healthy
subjects (second night of recording).  While the $\alpha$ values are
fluctuating, for REM sleep the $\alpha$ is larger than the $\alpha$ for 
light sleep, which is larger than the $\alpha$
for deep sleep (the 3 arrows indicate the cases which are not ordered
in the same way as the majority).  The exponent values have been 
determined over the fitting ranges as described in the caption of 
Fig.~\protect\ref{fig:3}.}
\label{fig:4}}
\end{figure}


\begin{thebibliography} {50}
\bibitem{fn1}  For a different order $m$ of detrending we obtain a different
 fluctuation function $F(n)$.  By construction, $F(n)$ is only defined for 
 $n \ge m+2$.
\end{thebibliography}
\end{document}